\documentclass[]{aa}

\usepackage{graphicx}
\usepackage{pjournal}
\usepackage{mamd}
\usepackage{natbib}
\usepackage{amssymb}

\begin{document}

\title{High resolution 21 cm mapping of the Ursa Major Galactic cirrus:
power spectra of the high-latitude \hi gas.}

 \author{M.-A. Miville-Desch\^enes\inst{1,}\inst{2,} \and G. Joncas\inst{3}\and E. Falgarone\inst{4}\and 
F. Boulanger\inst{1}}
 \institute{Institut d'Astrophysique Spatiale, Universit\'e Paris-Sud, B\^at. 121, 91405, Orsay, France
\and Canadian Institute for Theoretical Astrophysics, 60 St-George st, Toronto, Ontario, M5S 3H8, Canada
  \and D\'epartement de physique, de g\'enie physique et d'optique, Observatoire du mont M\'egantic, Universit\'e Laval, 
 Sainte-Foy, Qu\'ebec, G1K 7P4, Canada
\and LERMA/LRA, CNRS-UMR 8112, \'Ecole Normale Sup\'erieure, 24 rue Lhomond, 75005, Paris,
France}

\offprints{Marc-Antoine Miville-Desch\^enes}
\mail{mamd@ias.u-psud.fr}
\date{\today}

\titlerunning{Power spectrum of cirrus \hi emission}

\abstract{We present a power spectrum analysis of interferometric 21 cm observations of the 
Ursa Major high-latitude cirrus, obtained with the Dominion Radio Astrophysical Observatory (DRAO)
of Penticton (Canada). These high-resolution data reveal the intricate structure
of the diffuse Galactic \hip, at angular scales from 1 arcminute to 3 degrees. A filtering method
based on a wavelet decomposition was used to enhance the signal-to-noise ratio of the data.
The power spectra of the integrated emission and of the centroid velocity fields were used
to deduce the three-dimensional (3D) spectral index of the density and velocity fields of the \hi cirrus.
The spectral index is similar for the 3D density and velocity fields with a value of $-3.6\pm0.2$. 
Using the Leiden/Dwingeloo observations, this analysis was 
extended to the whole  North Celestial Loop (which includes the Ursa Major cirrus), 
showing that  the scaling laws prevail from 0.1 to 25 pc. 
The centroid velocity and integrated emission fields
show moderate correlation, with a maximum cross-correlation value of 0.44.}

 \maketitle


\section{Introduction}


It has been noted long ago \cite[]{chandrasekhar49,vonweizsacker51} that the interstellar medium (ISM) has 
the dynamical properties of a turbulent flow. The high Reynolds number of the ISM has often been invoked to ascribe 
its self-similarity, observed with several tracers on scales from a few tenths 
to several hundreds of parsecs, to the presence of a fully developed turbulence.
Nevertheless, despite the numerous efforts done to determine the dynamical properties of interstellar gas, 
the properties of interstellar turbulence are still poorly known.
This is mainly due to the difficulty to deduce three-dimensional properties of the gas
using projected quantities, but also to opacity effects and to the limited range of scales observed.
Characterizing interstellar turbulence, by determining its statistical properties for instance,
is still of major importance for the understanding of the structure and the kinematics of the ISM 
and its impact on the star formation process. 
In particular the statistical properties of the ISM structure and kinematics
can be used to identify regions with exceptional dynamical properties
and study their impact on the general evolution of the interstellar matter (see \cite{falgarone90,falgarone95a,joulain98}).
For instance this is essential to characterize the dynamical conditions that favor the cooling 
and condensation of \hi into its molecular form.
The same is true for the fragmentation/coagulation of dust grains \cite[]{falgarone95}.
On the theoretical side, there is a need for observational constraints for the ISM statistical properties.
These, like the power spectrum of density and velocity fluctuations in three dimensions (3D),
would nicely complement the numerical simulations efforts conducted lately (e.g. \cite{porter94}, 
 \cite{stone98},  \cite{mac_low99}, \cite{padoan99a}, \cite{lazarian2001}, \cite{heitsch2001}, \cite{ostriker2001}).

The statistical analysis of the information contained in line profiles is the natural venue of such study.
Regarding the column density fluctuations, several studies have shown the self-similar properties
of the molecular \cite[and references therein]{falgarone91,bensch2001} and dust emissions \cite[]{gautier92}
in several regions. For \hip, the power spectrum of the 21 cm emission has only been
characterized in external galaxies \cite[]{stanimirovic99,elmegreen2001,stanimirovic2001} and 
in the Galactic plane \cite[]{crovisier83,green93,dickey2001}.
These galactic studies were not centered on any particular \hi feature. Their analysis covered a depth in
velocity which sampled more than one spiral arm, leaving the necessity of studying isolated features.
Furthermore, the statistical properties of the column density do not allow to fully characterize
interstellar turbulence. For incompressible turbulent flows it is the velocity fluctuations power spectrum that
follows a power law. Density fluctuations will appear self-similar in response to the velocity field,
in certain conditions of compressibility or if the observed tracer is a passive scalar. Therefore, to 
study interstellar turbulence, it is important to deduce the statistical properties of both the velocity 
and density fields.

Centroid velocity fields have been used to study the kinematics of \hii regions 
\cite[]{joncas86,odell86,miville-deschenes95,godbout97} and to describe the velocity structure of 
star forming regions (see \cite{miesch99} and references therein) in which it is believed that the 
velocity field may be strongly coupled to the density field and could have a direct impact on 
the star forming process and the initial mass function \cite[]{scalo99}. 
For the \hi gas, there has been only works done on external galaxies \cite[]{spicker88,stanimirovic2001}
but no observational study thriving to determine the 3D statistical properties of Galactic interstellar velocity fields.
Recently, using fractional Brownian motion simulations, \cite{miville-deschenes2003b} 
showed that the power spectrum of the 3D velocity field of optically thin media can be determined 
directly from the centroid velocity field.

In this paper, we propose an analysis of interferometric 21 cm observations 
of the Ursa Major Galactic cirrus. The vast majority of the Galactic cirrus (named from their resemblance to
terrestrial cirrus in the IRAS all sky maps \cite[]{low84})
are composed of neutral hydrogen (\hi and H$_2$) and dust; they represent the most abundant state 
(by mass) of interstellar matter in the Galaxy. 
We have selected this nearby high latitude cloud for our statistical analysis to minimize the importance
of velocity crowding and to study the small scale structure of the \hip.
We have also chosen a cloud for which the inertial range of interstellar turbulence \cite[]{tennekes72} 
could be studied, meaning far from scales where energy inputs and dissipation occur (the viscous dissipation scale is 
thought to be $\sim$10 AU in the \hi - \cite{falgarone95}). 
The Ursa Major cirrus does not show any star formation activity \cite[]{magnani90} and it is
located at the edge of an expanding shell: the North Celestial Loop
\cite[]{heiles89,meyerdierks91}. Therefore, it is reasonable to assume that 
the main mechanical energy input comes from the
expanding shell, at scales much larger ($\geq$30 pc) than the field observed ($\sim$5 pc). The absence
of stellar winds or outflows prevents the presence of energy injection at intermediate scales.
Finally, the observed cloud is optically thin at 21 cm which is a required condition
to deduce the statistical properties of the 3D density and velocity fields from projected quantities
\cite[]{miville-deschenes2003b}.

The first three sections of the paper present the observations (\S~\ref{observations}), the processing used to 
enhance the signal-to-noise ratio of the data (\S~\ref{data_processing})
and the 21 channel maps (\S~\ref{data_ursa}). Then we present a power spectrum analysis 
of the integrated emission map, the centroid velocity field and of individual channel maps (\S~\ref{power_spectrum}). This
analysis allowed us to determine the spectral index\footnote{We refer to the spectral index as the 
exponent of the power spectrum assuming it can be represented by a power law.}  
of the 3D density and velocity fields of the \hi gas. Our results are discussed in \S~\ref{discussion}.


\begin{figure}[!t]
\includegraphics[width=\linewidth, draft=false]{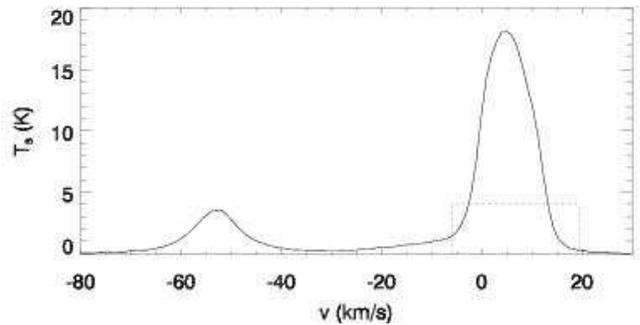}
\caption{\label{spectre_leiden} Low resolution average spectrum (from
the Leiden/Dwingeloo survey) of the Ursa Major cirrus observed with DRAO.
The spectral range covered at DRAO is delineated by a dotted line.}
\end{figure}

\section{Observations}

\label{observations}

The data were obtained using the Synthesis Radio Telescope of the Dominion Radio Astrophysical
Observatory (DRAO) \cite[]{landecker2000}. Two fields were observed and mosaiced. One in
1989 \cite[]{joncas92} and the second one in 1995. The observational parameters
of both fields are reproduced in Table~\ref{table_observ}. The only difference
between the two epochs was the addition of three new antennas to the interferometer.
In both cases the calibrators {were 3C147 and 3C309.1 (\cite{baars77}}. DRAO is also equipped with a 26m single dish telescope; 
{every HI observation taken with this dish is calibrated using  S7 (T$_{b}=$100 K)}. The
combination of the observations made by both telescopes (synthesis and single dish) allows the
complete coverage of the {\em u-v} plane down to the smallest scale resolved
by the synthesis telescope {(see \cite{taylor03} where the combination procedure is extensively described)}.

\begin{table}
\begin{center}
\begin{tabular}{ll}\hline
Parameter & Value \\ \hline \hline
Field 1 & $\alpha_{2000}= 9h47m$\\
& $\delta_{2000} = 70^\circ30'$\\
Field 2 & $\alpha_{2000}= 9h28m$\\
& $\delta_{2000} = 70^\circ10'$\\
Channel separation & 0.412 \kms \\
Channel resolution & 0.66 \kms \\
Number of channels & 128 \\
$V_{min}$ & -6.6 \kms \\
 $V_{max}$ & 19.4 \kms \\
Angular resolution & $0.99'\times 1.06'$\\
Field of view of each field & 2.6$^\circ$\\ \hline
\end{tabular}
\end{center}
\caption{\label{table_observ} Log of the DRAO observations. When the two fields are mosaiced together, this data set contains
a total of 107482 spectra.}
\end{table}

The mean 21 cm spectrum of the Ursa Major cirrus, 
taken with a beam of 30 arcminutes with the Leiden/Dwingeloo telescope \cite[]{burton94}, 
is shown in Fig~\ref{spectre_leiden}. 
The main \hi component is near 5~\kms but there is also
an intermediate velocity component near -55~\kmsp, possibly related to the Perseus arm.
In order to study the kinematics of the gas we have selected a channel width of 0.41~\kmsp.
The real velocity resolution is 0.65~\kms due to filter spillover.
Considering the number of channels of the spectrometer (see Table~\ref{table_observ})
we have limited our observations to the main \hi component (see Fig~\ref{spectre_leiden}).
This compromise between spectral resolution and velocity span allows us to
scan more than 80\% of the emission with a velocity resolution good enough to resolve
the thermal broadening of the CNM.

\begin{figure}[!t]
\begin{center}
\includegraphics[width=9cm, draft=false]{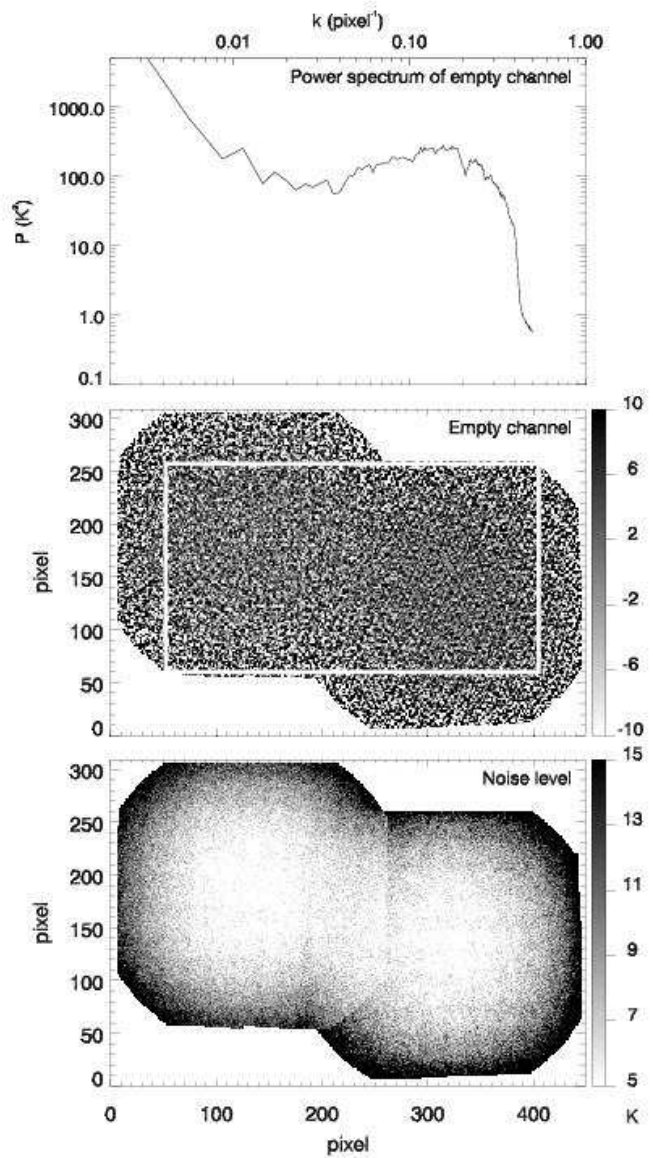}
\caption{\label{noise_charac} {\bf Middle:} Empty channel of the combined Ursa Major observations.
{\bf Top:} Power spectrum of a sub-region of the empty channel (white rectangle in
Middle panel). {\bf Bottom:} Noise level at each position,
estimated by the standard deviation of each spectrum high frequency fluctuations.}
\end{center}
\end{figure}

\section{Data processing}

\label{data_processing}

The preliminary steps of the reduction were done at DRAO 
(combination of the observations from the interferometer and from the
single dish, construction of the channel maps). These standard operations are 
described in \cite{joncas92}. 
The observations presented in this paper are characterized by a somewhat low signal-to-noise ratio 
($\lesssim 15$ - see Fig.~\ref{spectre_avant_apres}). 
This is partly compensated in the statistical studies we present
here by the large size of the sample. Once mosaiced together, the data set 
contains 107482 spectra\footnote{As the angular resolution is $\sim 1'\times 1'$ and
the pixel size is $0.5'\times 0.5'$, the number of independent spectra is 1/4 of that value 
($\sim 26 870$).}.
In this section we present the filtering method we have used to increase the signal-to-noise ratio of the data.

\begin{figure*}[!ht]
\includegraphics[width=\linewidth, draft=false]{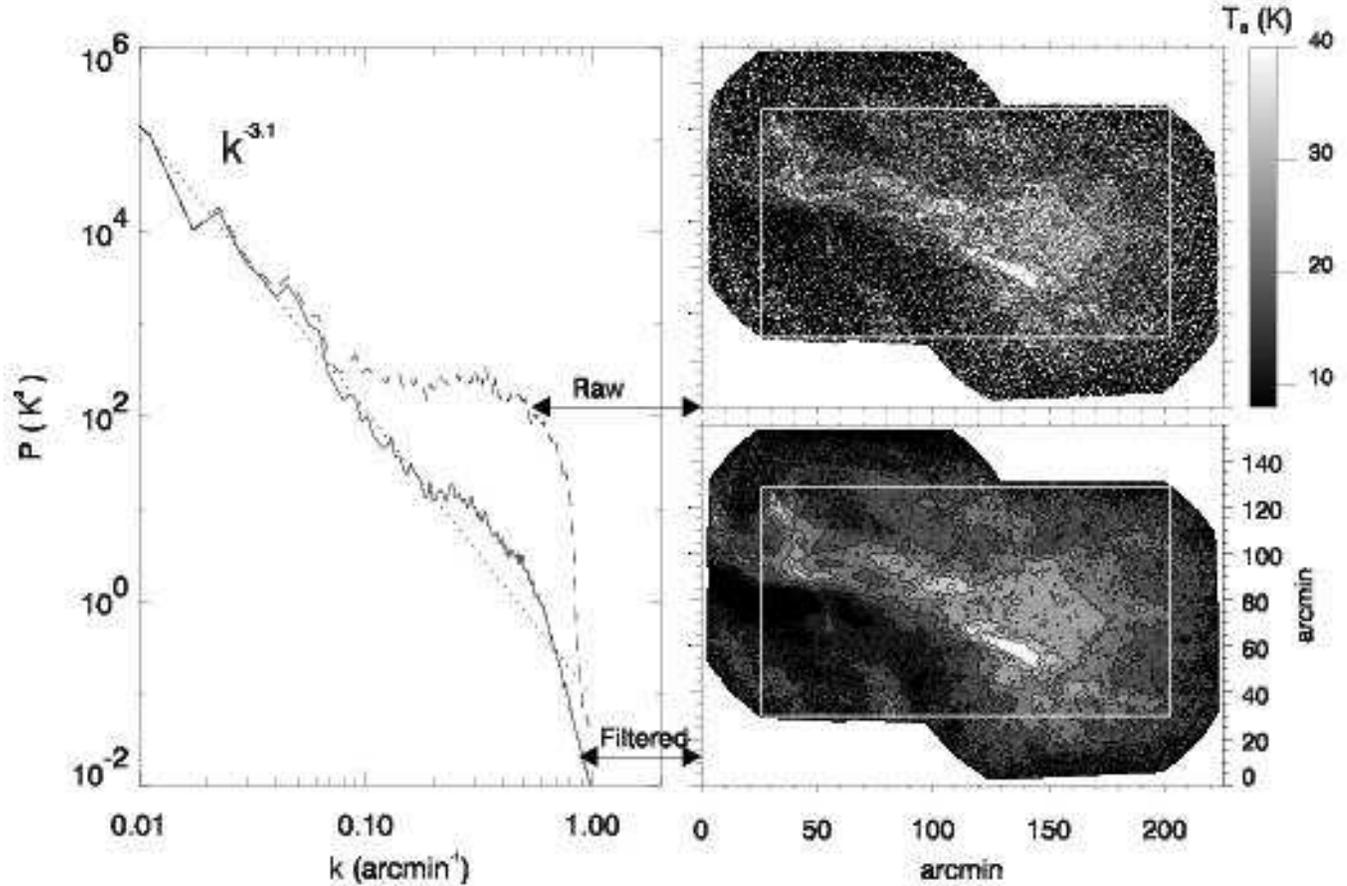}
\caption{\label{exemple_filtrage} Example of the filtering on an individual channel map. {\bf Top-right:}
raw data. {\bf Bottom-right:} filtered data. Note the presence 
of a weak elongated structure in the south-eastern part of the field,
running from (50', 80') to (90', 40'), barely visible in the raw data and preserved by the
filtering method. {\bf Left: } Power spectrum of the channel map 
(inside the white rectangle area), before and after filtering. The pixel size is $30''\times30''$.}
\end{figure*}

\subsection{Characterization of the noise}

The observations have been conducted such that there are some channels with basically
no \hi emission. These channels are used to characterize the noise. 
An empty channel of the combined Ursa Major observations (see Fig~\ref{noise_charac} - middle)
shows that the noise level is not spatially uniform.
In the combination of the interferometric and single dish observations we have normalized 
the data for the beam of the
single dish. This operation increases the noise level near the edge of the field.
This increase is quantified by computing the standard deviation of the
high frequency fluctuations (high-pass filtering) on each spectrum of the data cube (see Fig~\ref{noise_charac} - bottom). 
The noise level varies from $\sim 3$ to $\sim 13$ K from the center to the edge of the
field. {Fig~\ref{noise_charac}-top shows the power spectrum\footnote{The technique used
to compute the power spectrum is described in \S~\ref{integrated_emission}.} 
of a sub-region of an empty channel (see white rectangle
in Fig~\ref{noise_charac} - middle). 
There is an increase of power at large scales
($k < 0.02$ pixel$^{-1}$) mainly due to the increase of the noise level on the side of the field.
Part of this increase could also be attributed to the lower statistics of the large scale, 
an effect inherent to finite size images.
At small scales ($k > 0.3$ pixel$^{-1}$) the abrupt power drop is due to the
beam of the interferometer. But in the intermediate regime the power spectrum
is relatively flat which is an indication of the uniformity of the u-v plane coverage.}


\subsection{Filtering}

To increase the signal-to-noise ratio of this type of spectro-imaging
data we have developed a method 
based on a wavelet decomposition of each channel. This method, 
an adaptation of the work of \cite{starck98}, looks for
structures that are significant in terms of signal-to-noise ratio and
that are also coherent in the spectral domain. To optimize the filtering
we have used our knowledge of the spectral and spatial variations of the noise level.

First each channel $c_0(x,y,v)$ of the data cube was decomposed on a wavelet basis using the ``a trou'' algorithm:
\begin{equation}
\label{decomp}
c_0(x,y,v) = c_p(x,y,v) + \sum_{l=1}^p w_l(x,y,v),
\end{equation}
where the $w_l(x,y,v)$ are the wavelet coefficients at a given scale $l$ and the $c_p(x,y,v)$ are
the large scale residuals. Then, the wavelet coefficients $w_l(x,y,v)$
are compared to the amplitude of the noise $n_l(x,y)$ at each scale and position.
The $n_l(x,y)$ were computed as
\begin{equation}
n_l(x,y) = A_n(x,y) \times \sigma_n(l)
\end{equation}
where $A_n(x,y)$ is the noise level as a function of position (Fig.~\ref{noise_charac}-bottom)
and $\sigma_n(l)$ is the dispersion of the wavelet coefficients of a white noise image with a standard deviation of one
\cite[]{starck98}.

If the wavelet coefficient satisfies the following criterion:
\begin{equation}
|w_l(x,y,v)| > 1.5 \times n_l(x,y)
\end{equation}
it is kept, otherwise it is put to zero. Each channel is then reconstructed with
the significant wavelet coefficients using Eq.~\ref{decomp}.
\cite{starck98} suggest a threshold of 3 but we have chosen to be
more conservative using a threshold of 1.5 to limit the removal of real 
weak and extended structures from the data.

This filtering removes all the fluctuations that do not have a significant 
signal-to-noise ratio, but there are always strong noise peaks that are not removed. 
To detect these noise peaks, we use the fact that they
are not correlated from one channel to the other. 
In practice, each spectrum is checked for discontinuities larger than 8 K
that go up and down (or down and up) in one or two channels. These noise peaks
were removed and the missing values were interpolated on the spectral axis.

\begin{figure}
\hspace{-0.8cm}
\includegraphics[width=10cm, draft=false]{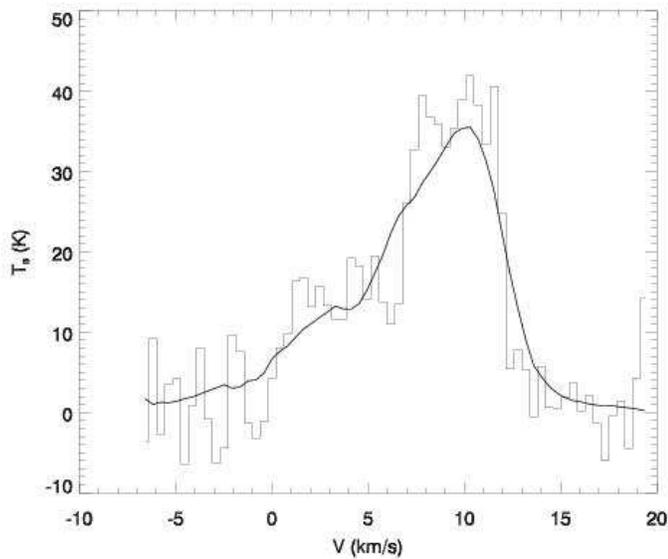}
\caption{\label{spectre_avant_apres} Typical 21 cm spectrum of the DRAO observation, 
before (histogram) and after (continuous line) the filtering.}
\end{figure}

An example of the filtering on a channel map is shown in Fig~\ref{exemple_filtrage}.
{The power spectrum of this channel map, computed before and after filtering
(see Fig~\ref{exemple_filtrage}) shows that the noise in the raw data dominates
the signal at scales smaller than $\sim 20'$. With the filtering technique applied here,
we were able to lower the noise significantly, decreasing the scale range dominated
by noise. In the filtered channel map shown here, the noise becomes important at scales
smaller than $\sim 4'$ (8 pixels).}

The filtering applied to the data allows the detection of spatially coherent structures that had a signal-to-noise
ratio down to 1 in the raw channel maps. For instance, in Fig~\ref{exemple_filtrage} we note the presence of a
weak elongated structure in the south-eastern part of the field, running from (50', 80') to (90', 40'),
that was barely visible in the raw data. This very thin
filament has a thickness of $\sim 0.06$ pc and a length of greater than 1 pc.
The filtering has also reduced considerably the noise on individual spectra 
as shown in Fig.~\ref{spectre_avant_apres}. 
This is of major importance for the analysis of the velocity field.

\begin{figure*}
\begin{center}
\includegraphics[width=18cm, draft=false]{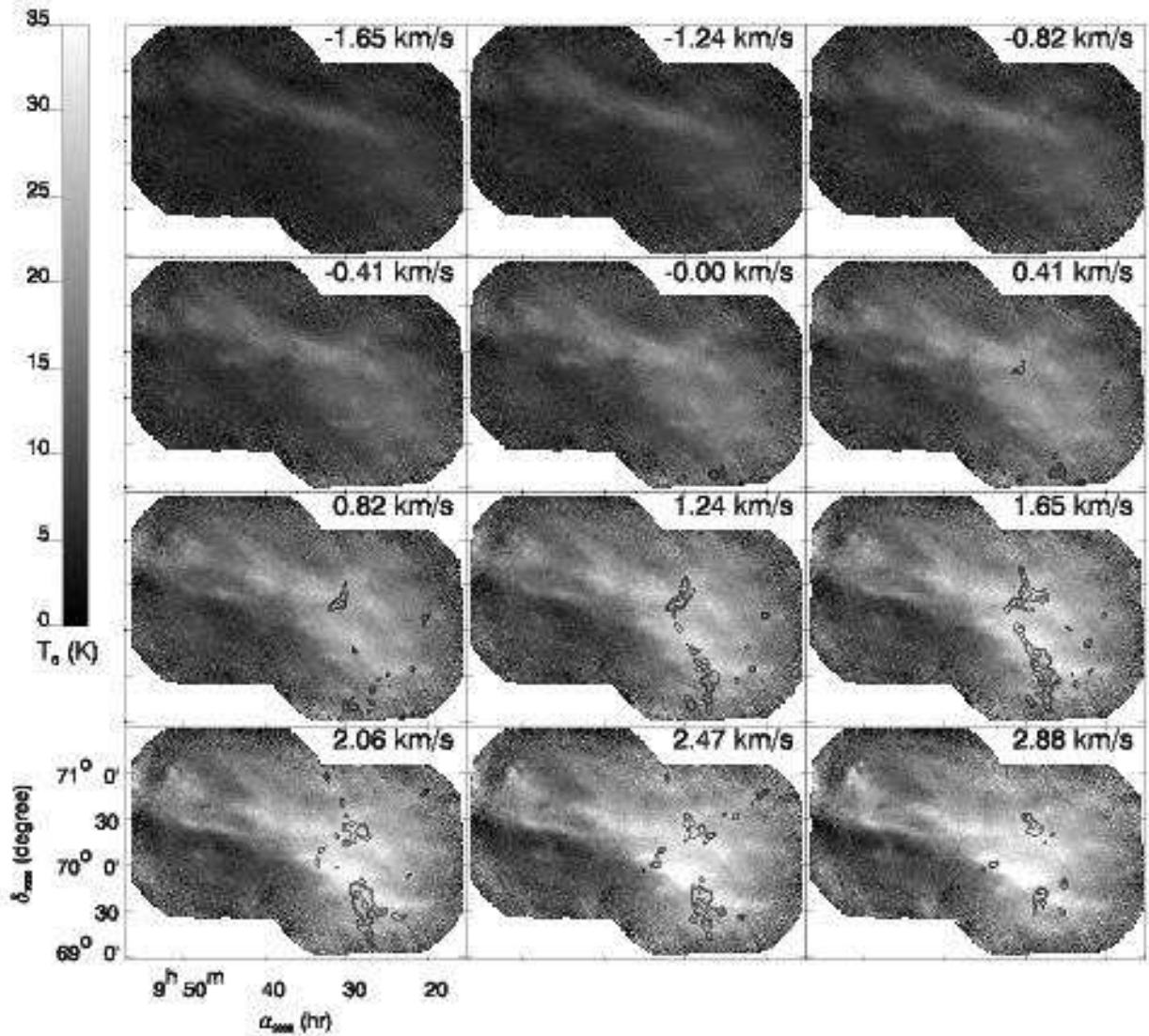}
\caption{\label{carte_ursa2} Ursa Major. Channel maps of the 21 cm line emission between -1.65 and 2.88 \kmsp.
The contours (1.5, 2 and 3 K) represent the CO emission of \cite{pound97}.}
\end{center}
\end{figure*}

\begin{figure*}
\begin{center}
\includegraphics[width=18cm, draft=false]{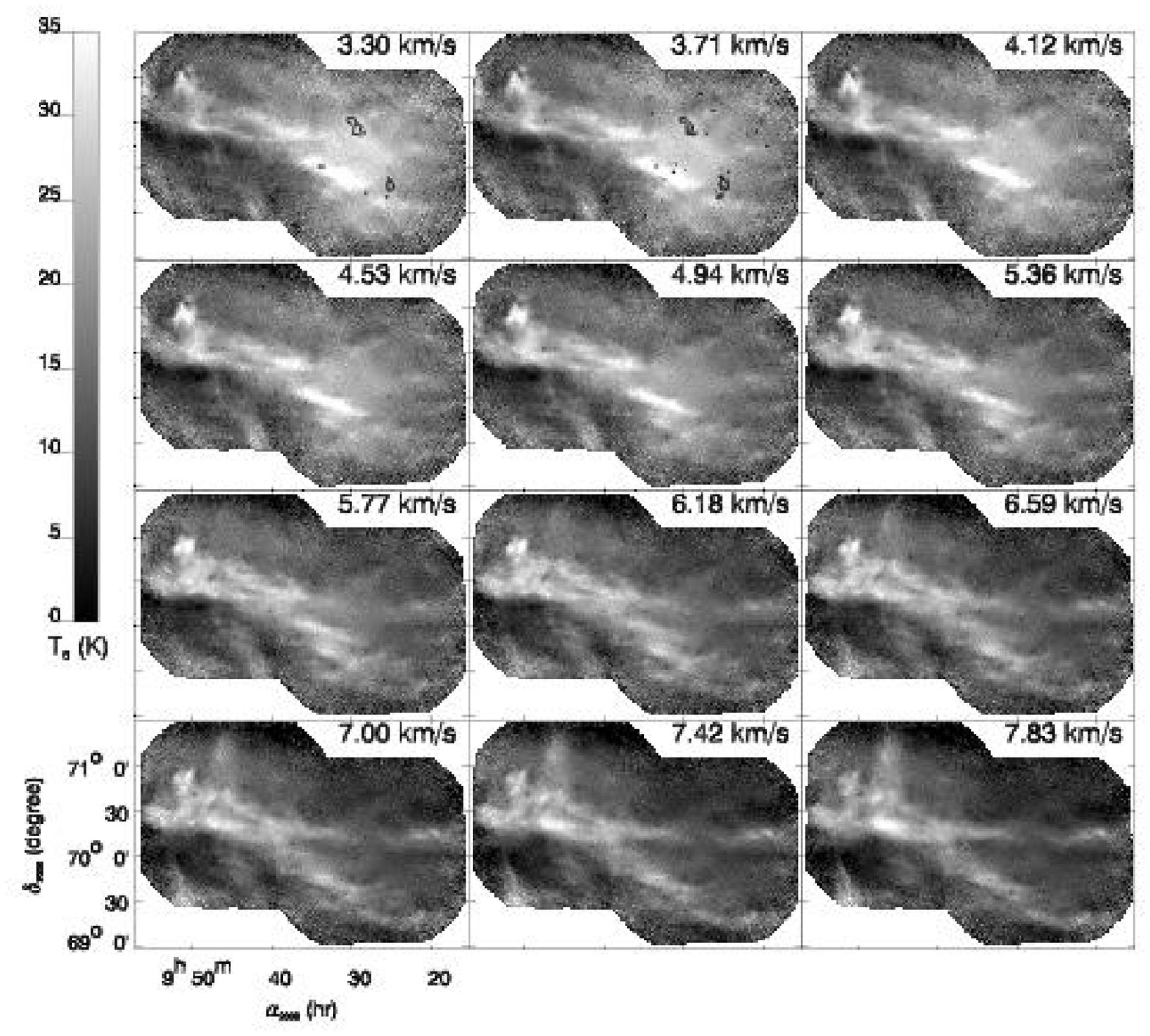}
\caption{\label{carte_ursa1} Ursa Major. Channel maps of the 21 cm line emission between 3.30 and 7.83 \kmsp.
The contours (1.5, 2 and 3 K) represent the CO emission of \cite{pound97}.}
\end{center}
\end{figure*}

\begin{figure*}
\begin{center}
\includegraphics[width=18cm, draft=false]{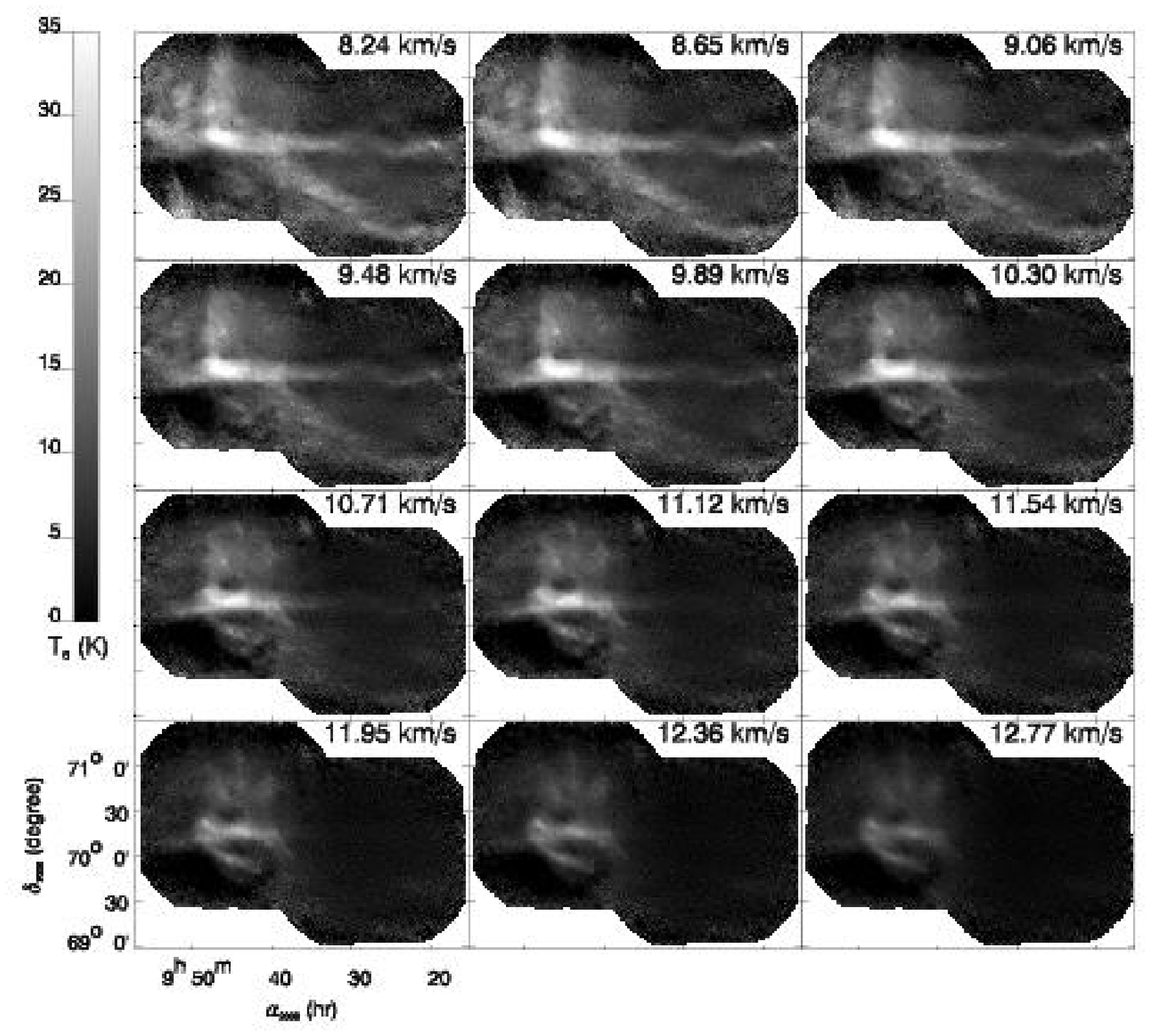}
\caption{\label{carte_ursa0} Ursa Major. Channel maps of the 21 cm line emission between 8.24 and 12.77 \kmsp.}
\end{center}
\end{figure*}

\section{The 21 cm data of the Ursa Major cirrus}

\label{data_ursa}

\subsection{21 cm channel maps}

The Ursa Major cirrus is composed of two connected parts. The eastern region (field 1) is predominantly atomic
as it is detected only in the \hi line. The western region (field 2) presents significant CO emission
\cite[]{pound97}. The data from field 1 has already been discussed by \cite{joncas92} but the data processing techniques 
used since then have given a second life to these data. Figures~\ref{carte_ursa2} to \ref{carte_ursa0} 
show the \hi mosaic in a series of panels separated by 0.412 \kms. The velocity is indicated in the upper
right portion of each panel. The brightness temperature in the channel maps varies from 1 K to 40 K. 
The CO emission as observed by \cite{pound97} is present from 0.0 \kms to 3.71 \kms. 
It is shown in contours, superimposed on the \hi emission, in Figures~\ref{carte_ursa2} to \ref{carte_ursa0}.

The 21 cm channel maps show an intricate structure of interwoven elongated structures
{with aspect ratio up to 15}. For the sake of simplicity, we call them filaments in the following,
but, as indicated by \cite{lazarian2000}, we are aware that the structure identified in 
any channel map reflect more velocity features than density enhancements. 
{Therefore we deliberately restrain the description of the structure observed in channel maps.
Nevertheless it is interesting to point out that the western region, where CO emission is observed, shows generally 
less filamentary structure than the eastern part of the field where no molecular emission is detected.
The detailed comparison of the molecular and atomic components of the cloud, which is
clearly beyond the scope of this paper, is important to understand the specific dynamical conditions 
that leads to the formation of long-lived molecular clouds. We leave it to a further study.}


\subsection{Integrated emission map}

One important quantity that can be derived from 21 cm data cubes is the
integrated emission which, for optically thin media, is related to
the column density of \hip. 
The 21 cm brightness temperature $T_B(v)$ at a given velocity $v$ and within $\delta v$ of \hi gas having 
a kinetic temperature $T_k$ 
is given by: 
\begin{equation}
T_B(v)  = T_k \left( 1 - \exp\left(-\tau(v)\right) \right)
\end{equation}
where
\begin{equation}
\tau(v) = \frac{N(v)}{C\times T_k \delta v},
\end{equation}
$N(v)$ being the gas column density at velocity $v$ and within $\delta v$, and
$C = 1.823\times10^{18}$ cm$^{-2}$ K$^{-1}$ km$^{-1}$ s  \cite[]{kulkarni87}.
In the optically thin regime ($\tau(v) < 1$), the column density can be estimated
without knowing the kinetic temperature of the gas. In the optically thin approximation
\begin{equation}
\label{eq_ntot}
N(v) = C\times T_B(v) \delta v.
\end{equation}

\begin{figure}[!ht]
\begin{center}
\includegraphics[width=\linewidth, draft=false]{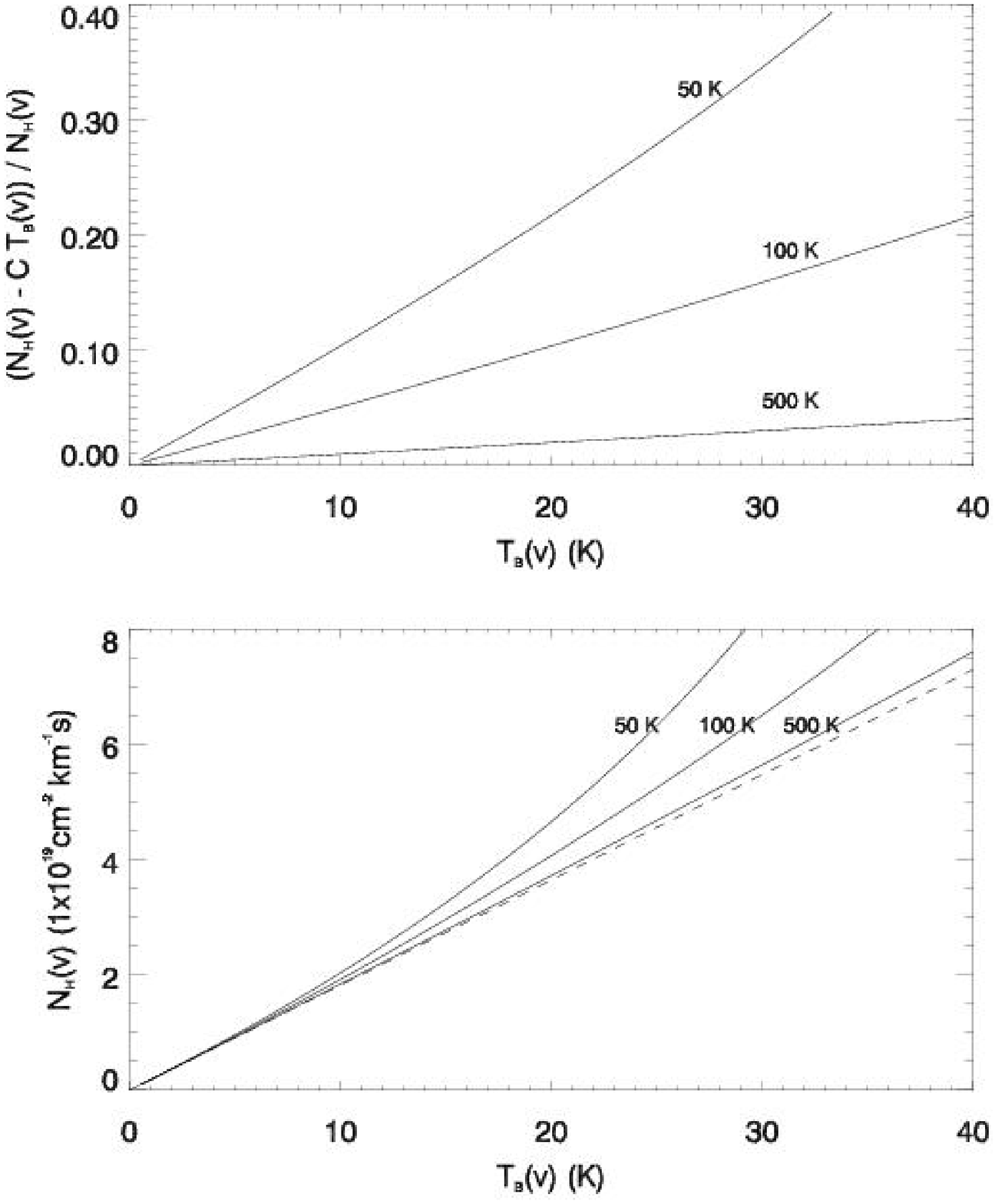}
\caption{\label{tau_and_temperature} Error done on the column density estimate
using the optically thin approximation (Eq.~\ref{eq_ntot}) for \hi gas with a kinetic temperature of 50, 100 and 500 K. 
{\bf Bottom:}  The column density (per velocity interval) determined using the optically thin approximation as a 
function of brightness temperature
is given by the dashed line. The three continuous lines indicate the real column density (per velocity interval) 
of the gas as a function of brightness temperature.
{\bf Top:} Fraction of the real column density that is
missed in the optically thin approximation as a function of the brightness temperature at a given velocity.}
\end{center}
\end{figure}

As shown in Fig.~\ref{tau_and_temperature}, the effect of opacity on the determination of the
column density is stronger for cold gas and for high brightness temperatures. 
For warm \hi ($T_k > 500 $K), the opacity is negligible and the column density is well
approximated by Eq.~\ref{eq_ntot}. On the other hand, for typical CNM kinetic temperature ($\sim$100 K)
and for brightness temperature of $\sim$40 K (the maximum value in our channel maps),
the column density in a given channel map is underestimated by $\sim$20\%. But the 21 cm profiles
are relatively broad and most of the brightness temperature values are below 20 K. Therefore, if the kinetic
temperature of the \hi in Ursa Major is $\sim$100 K, we think that the column density is probably 
underestimated by only $\sim10$\% in the brightest region of the cloud, which confirms that we
are in the optically thin regime.

Instrumental noise also affects the determination of the column density.
Since the uncertainty on the column density increases with 
the square root of the number of independent channels used in the summation,
it is common to define a spectral window within which the column density 
is computed. This way noise contribution from the empty channels is avoided.
However for these observations the baseline is very short (see Fig.~\ref{spectre_avant_apres})
and varies in length at both ends as a function of the Doppler shifts of the line emission. Therefore all 
channels were summed. 

The integrated emission map of the Ursa Major field is shown in  Fig.~\ref{ps_ntot_ursa}.
This map is characterized by a set of filaments, mainly oriented west-east. 
The integrated emission peaks at $W \approx 330$ K \kms (corresponding to $N_H\sim 6\times 10^{20}$ \ccap)
in the eastern part of the main filament. The mean integrated emission value in the field is $W \approx 190$ K \kms
($N_H\sim 3.45\times 10^{20}$ \ccap).

\subsection{Centroid velocity field}

The centroid velocity field $C(\alpha, \delta)$ of a position-position-velocity (PPV) data 
cube $T_B(\alpha, \delta, v)$ is given by the following expression:
\begin{equation}
C(\alpha, \delta) = \frac{\sum_v v \times T_B(\alpha,\delta, v)}
        {\sum_v T_B(\alpha,\delta,v)}
        \, \mbox{ \kms}.
\end{equation}
For optically thin clouds, the centroid velocity field is exactly the mean radial velocity of
the gas on the line of sight. 
As shown by \cite{pety99}, noise significantly affects the determination of
velocity centroids. In fact, the error on the velocity centroid determination depends
on the signal-to-noise ratio of the spectrum and on the number of channels used. 
To minimize the effect of noise, one may compute the centroid velocity on each spectrum 
by selecting a spectral window on each spectrum where all the brightness temperatures
are greater than twice the noise level.
But practically, the wavelet filtering we have applied 
significantly reduces the noise level of the data leaving very few
noise peaks that could contaminate the centroid velocity estimate. 
Therefore, and as there are very few channels with no emission, all the channels were 
used to compute the centroid velocity of each spectra. The centroid velocity field of 
the Ursa Major cirrus is shown in Fig.~\ref{ps_velfield_ursa}.

{The centroid velocity field of the Ursa Major cirrus is characterized
by a set of elongated structures. 
In fact, the two main elongated structures seen in the integrated emission map and
in channel maps also stand out in the centroid velocity field. Velocity fluctuations
are observed at all scales, up to the largest one where there is
an east-west velocity gradient of $\sim 1.5$~km~s$^{-1}$~pc$^{-1}$ at a 2$^\circ$ scale
(considering a distance of 100 pc for the cirrus). 
But it is at small scales that the velocity gradients are the largest.
Across the brightest filament of the cirrus (near position (50', 80') in Fig.~\ref{ps_velfield_ursa})
we observe transverse velocity gradients between $\sim 5$ and 10~km~s$^{-1}$~pc$^{-1}$, at a 20' scale
(also shown by \cite{miville-deschenes2002}).}


\section{Power spectrum analysis}

\label{power_spectrum}

\subsection{Determination of the density and velocity power spectra in three dimensions}

{One important aspect of the study of interstellar turbulence is to determine
the statistical properties of the density and velocity fields in three dimensions (3D).
To do so one must understand the effect of opacity and of projection on the observed quantities.
Regarding density fluctuations, many authors \cite[]{stutzki98,lazarian2000,goldman2000,miville-deschenes2003b}
showed that the power spectrum of the 3D density field can be determined
directly from the power spectrum of the integrated emission map under two conditions: 
1) the observed medium must be optically thin and 2) the spatial scales observed on 
the plane of the sky must be smaller or equal to the spatial depth of the line of sight.
Under these conditions, the power spectrum of the integrated emission map
is proportional\footnote{It has the same shape - or spectral index - but not the same normalisation.} 
to the power spectrum of the 3D density field. }

{The determination of the 3D velocity field power spectrum is less obvious.
\cite{lazarian2000} proposed a method called Velocity Channel Analysis (VCA)
that uses the variation of the spectral index of velocity slices of increasing thickness.
But as pointed out by \cite{miville-deschenes2003b} and as we will show here, this 
method is very difficult to apply on real observations. On the other hand,  \cite{miville-deschenes2003b}
showed that centroid velocity fields can be used to determine the power spectrum of the 3D velocity field.
Using Fractional Brownian Motion simulations\footnote{Fractional Brownian motion simulations 
produce self-similar Gaussian random images.} these authors 
studied the impact of density and velocity fluctuations on the 
power spectrum of centroid velocity fields. They showed that density fluctuations
have a very limited effect and that the power spectrum of the centroid velocity field 
is the same as for the 3D velocity field, assuming that the medium observed is optically 
thin and that the scales observed on the sky are smaller than the depth of the line of sight. }

{It is important to estimate if these results can be applied to the
specific physical conditions of the Ursa Major cirrus.
First, we have shown that opacity effects are limited for the brightness temperature observed
here (see Fig.~\ref{tau_and_temperature}). 
Second, as the cloud is at 100 pc, the largest transverse scale in the field
is only $\sim$7~pc, which is very likely to be smaller than the depth of the \hi froth in that
region (probably several tens of pc if we look at large scale observations - see Fig.~\ref{ncl_leiden}). 
Therefore, in the following we analyze the power spectrum of the integrated emission map 
and of the centroid velocity field of the Ursa Major cirrus in the framework of \cite{miville-deschenes2003b}.}



\begin{figure*}[!ht]
\hspace{-0.7cm}
\includegraphics[width=\linewidth, draft=false]{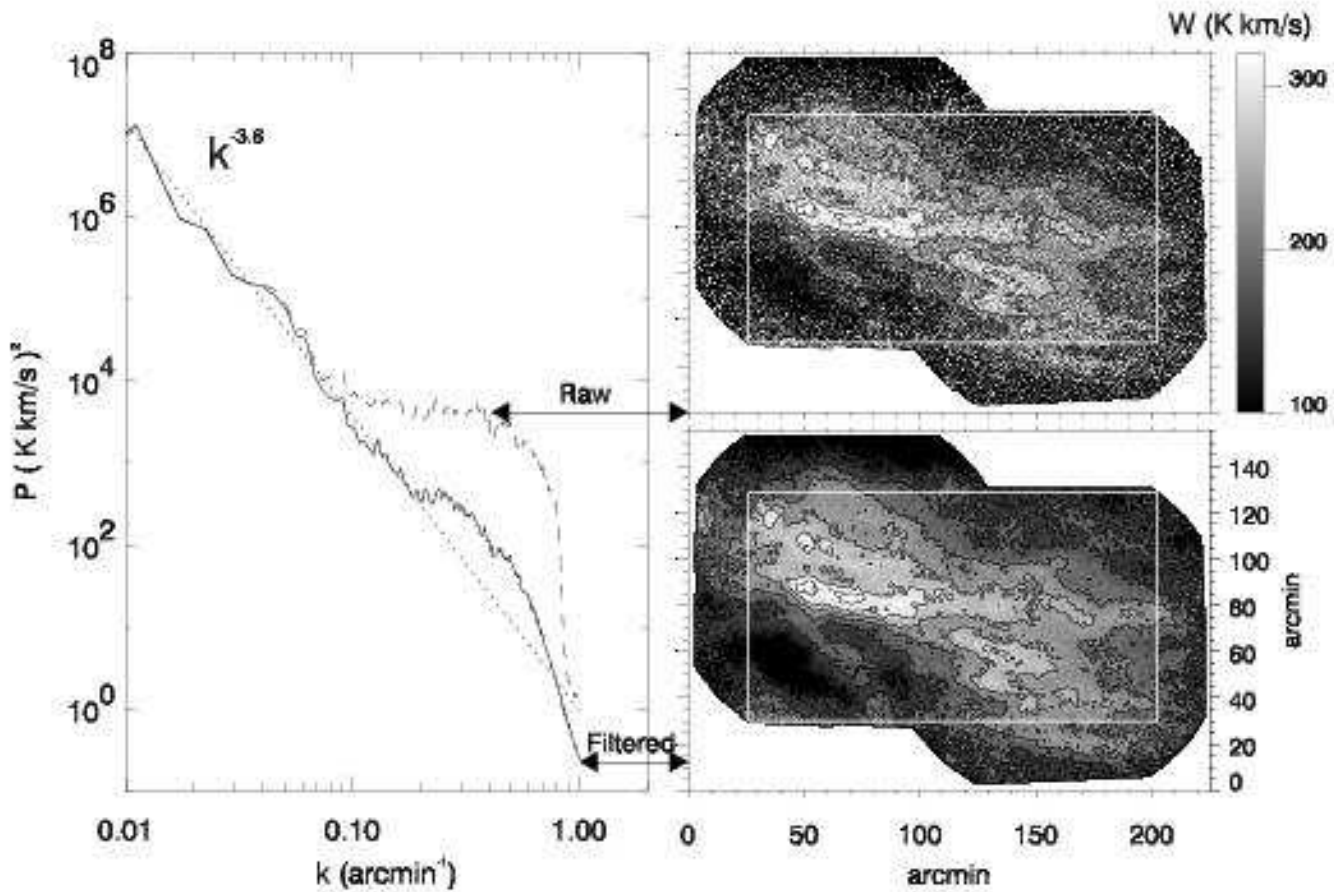}
\caption{\label{ps_ntot_ursa} \hi Integrated emission of the Ursa Major cirrus {\bf Right-top:} raw data. 
{\bf Right-bottom:} filtered data. {\bf Left:} Power spectrum of the integrated emission map
of the raw and filtered data. The dotted line represents a power law with a spectral index of -3.6.}
\end{figure*}

\subsection{Integrated emission map and centroid velocity field}

\label{integrated_emission}

\begin{figure*}[!t]
\hspace{-0.7cm}
\includegraphics[width=\linewidth, draft=false]{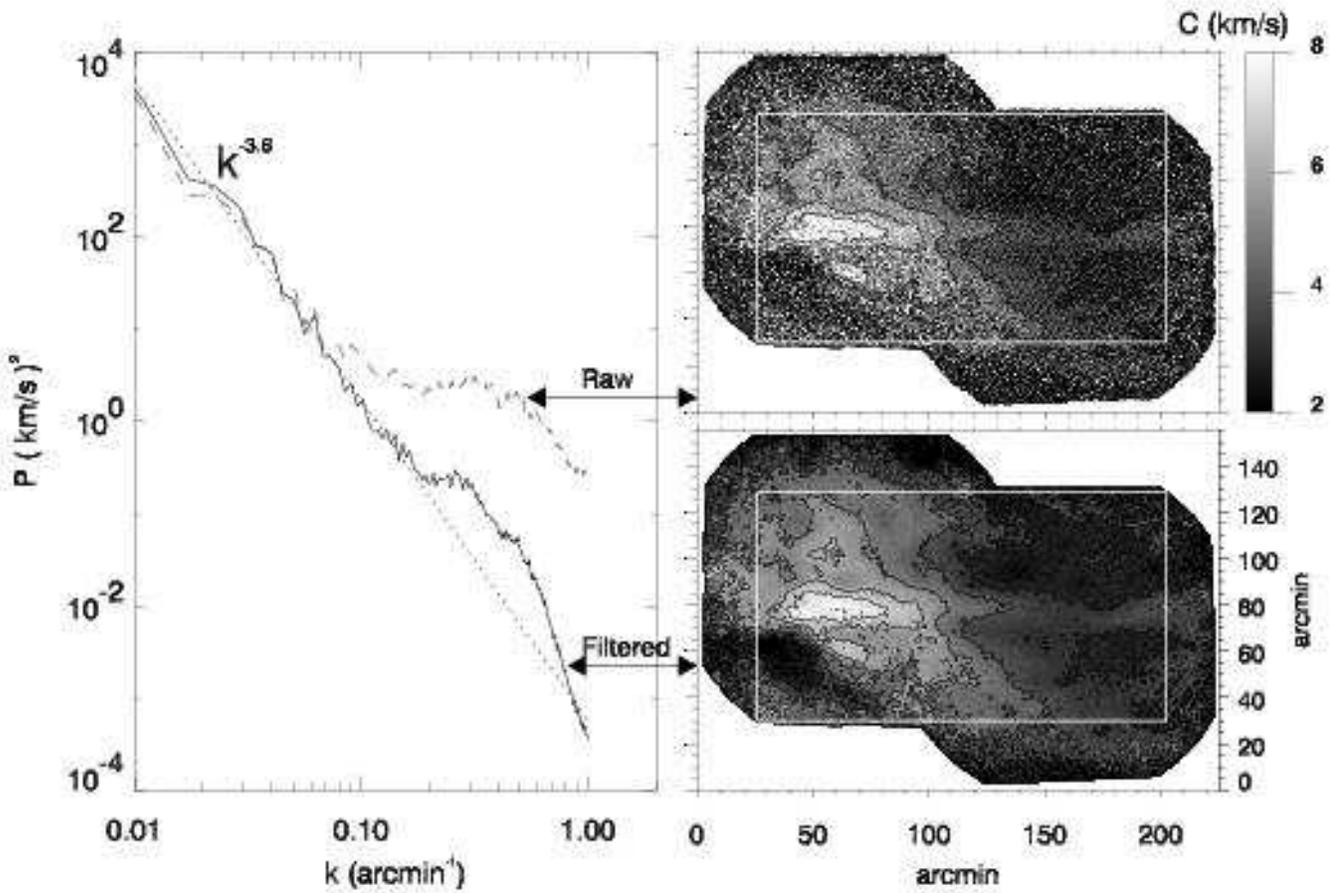}
\caption{\label{ps_velfield_ursa} \hi centroid velocity field of the Ursa Major cirrus {\bf Right-top:} raw data. 
{\bf Right-bottom:} filtered data. {\bf Left:} Power spectrum of the centroid velocity field
of the raw and filtered data. The dotted line represents a power law with a spectral index of -3.6.}
\end{figure*}

{For the Ursa Major observations, we have limited the power spectrum computation 
to a rectangular region ($354 \times 198$ pixels) where all the points have been observed.
This sub-region is delimited by a white rectangle in 
figures~\ref{exemple_filtrage}, \ref{ps_ntot_ursa} and \ref{ps_velfield_ursa}.
The power spectrum is the sum of the square of the real and imaginary
parts of the two dimensional FFT, azimuthally averaged over each wavenumber $k=\sqrt{k_x^2 + k_y^2}$.
As stated by \cite{bensch2001}, edge effects may introduce systematic bias in the
determination of the power spectrum index using Fast Fourier Transform algorithms.
When the power spectrum is described by a power law $P(k) \propto k^{\beta}$, 
the error on the power spectrum index depends on $\beta$ itself; it becomes important
for $\beta < -3$. To reduce this effect, we have apodised the edge of the image
with a cosine function. The size of the apodised region is equal to 3\% of the linear size
of the image. We have tested this method on non-periodic Fractional Brownian Motion images
\cite[]{miville-deschenes2003b} and we have found that the $\beta$ value is recovered with 
a typical error of 0.1 for $-4<\beta<-2$.}

{The power spectra (raw and filtered data) of the integrated emission map and centroid velocity field
are shown in Fig.~\ref{ps_ntot_ursa} and \ref{ps_velfield_ursa}. 
For both the integrated emission map and the centroid velocity field
the power spectrum of the filtered data is well described by a power law $\propto k^{-3.6 \pm 0.2}$
for wavenumbers $0.01 < k < 0.2$ arcmin$^{-1}$. 
We have to restrict the slope determination to a relatively narrow scale range for two reasons.
First, at large scale, the 1D power spectrum points have a large uncertainty as they are the average
of a small number of samples in the 2D FFT. Second, at small scales, the power spectrum is affected by the
beam, the noise and the non-periodicity of the image mentioned earlier. 
The uncertainty on the spectral index we give in this paper ($\pm0.2$) reflects the dispersion of the $P_k$ values 
around the power law but mainly the uncertainty related to the apodization, beam and noise (see above). 
This uncertainty value has been validated using the fractional Brownian motion simulations and realistic noise and 
beam shape.}

{As opacity effects are negligeable at 21 cm and the depth of the cirrus is most probably 
larger than the largest transverse scale, we estimate that the 3D density and velocity fields of 
the Ursa Major cirrus are characterized by a spectral index $\gamma_v = \gamma_n = -3.6 \pm 0.2$.
Here we consider that the energy driving the turbulent motions is injected at much larger scales
than the size of the observed fields and that the behavior of the velocity field at all scales is 
inherent to the turbulent cascade. 
It is interesting to point out that the velocity spectral index is compatible with 
the value expected (-11/3) for Kolmogorov turbulence \cite[]{kolmogorov41}. }

\subsection{Velocity channel analysis}

\begin{figure}[!t]
\hspace{-0.7cm}
\includegraphics[width=9cm, draft=false]{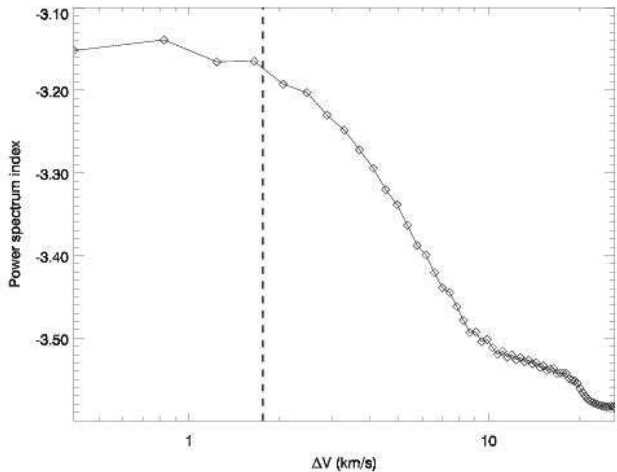}
\caption{\label{expo_vs_channel} Power spectrum index of velocity slices as a function
of the slice width $\Delta V$. Velocity slices are built by adding velocity channel maps 
from the the brightest to the faintest.}
\end{figure}

{As discussed earlier, \cite{lazarian2000} proposed a method to determine the spectral index
of the 3D density and velocity fields using PPV data cubes of optically thin media.
This method has been used by a few authors
to determine the 3D velocity spectral index $\gamma_v$ of \hi in the Galactic plane \cite[]{dickey2001}
or in other galaxies  \cite[]{stanimirovic99,stanimirovic2001,elmegreen2001}. 
The VCA method is based on the analysis of velocity slices of increasing width $\Delta V$. 
\cite{lazarian2000} claim that the spectral index of velocity slices should reach two asymptotic states 
at small and large $\Delta V$, from which $\gamma_v$ could be determined.

We have tested the VCA method on our observations. Fig.~\ref{expo_vs_channel}
shows the variation of the spectral index of velocity slices of increasing width $\Delta V$.
Here a velocity slice of width $\Delta V$ is the sum of the $N=\Delta V/\delta u$ brightest consecutive
channel maps, where $\delta u=0.412$ \kmsp. The spectral index of each velocity slice
was computed; it is the slope of the power spectrum between $0.01 < k < 0.1$ arcmin$^{-1}$. 
As expected, at large $\Delta V$ the curve in Fig.~\ref{expo_vs_channel} converges to the spectral index
of the integrated emission ($\sim$-3.6). At the other end, the curve reaches another asymptotic state 
to the value ($\sim$-3.1) found for the brightnest channel map (see Fig.~\ref{exemple_filtrage}).

But here it is important to estimate the effect of the gas temperature on this curve.
The vertical dashed line in Fig.~\ref{expo_vs_channel} shows the effective
spectral resolution of our observations $\delta v_{eff} \approx \sqrt{\delta u^2 + 2.16k_B T /m } = 1.7$~\kms 
(where $m$ is the mean atomic mass and if we consider typical 
CNM gas at $T=150$ K - see \cite{miville-deschenes2003b} for details).
The plateau seen at small $\Delta V$ in Fig.~\ref{expo_vs_channel} is simply
the result of the gas temperature that washed out all velocity fluctuations in
channel maps. Therefore, the asymptotic state at small $\Delta V$ is not the one expected
by \cite{lazarian2000} and cannot be used to determine $\gamma_v$.
Like \cite{miville-deschenes2003b} we conclude that 
while it could be used for large scale observations (i.e. external galaxies), 
the VCA technique is not suited to determine the 3D velocity spectral index
for local HI where the velocity dispersion is commensurate with the thermal
broadening on a broad range of scales.}


\subsection{Combination with Leiden observations}

\label{section_leiden}

To obtain a more global view of the \hi power spectra (integrated emission and centroid velocity) 
we combined the power spectra computed on the Ursa Major DRAO observations 
with the power spectra computed on the Leiden/Dwingeloo observations of
the whole North Celestial Loop region (which includes the Ursa Major cirrus)
as observed by the Leiden/Dwingeloo survey (see Figures~\ref{ncl_leiden} and \ref{velfield_ncl}). 
The angular scales of each observation complement very well to give the power
spectrum of the \hi integrated emission and of the centroid velocity fields 
on almost three orders of magnitudes in scales, from 2 arcminutes to 16 degrees (see Figures~\ref{ps_ntot_leiden} and \ref{ps_velfield_leiden}). 
One outstanding result of this analysis is the fact that both power spectra 
are compatible with a single slope of $-3.6\pm0.2$ over the whole range of scales. The decrease of power at small scales
($k>0.4$ arcmin$^{-1}$) is due to the domination of the noise at these scales.
All the power at these scales was removed by the wavelet filtering. At the other end of the power spectrum, 
the flattening may not be significant as the statistics at large scale are low.
If we consider that most of the gas in the North Celestial Loop is local, 
at an approximate distance of 100 pc \cite[]{penprase93}, and the opacity effects are limited
in the region, the local \hi is self-similar on spatial scales from 0.5 to 25 pc.

\subsection{Density-velocity correlation}

\begin{figure}[!t]
\includegraphics[width=9cm,draft=false]{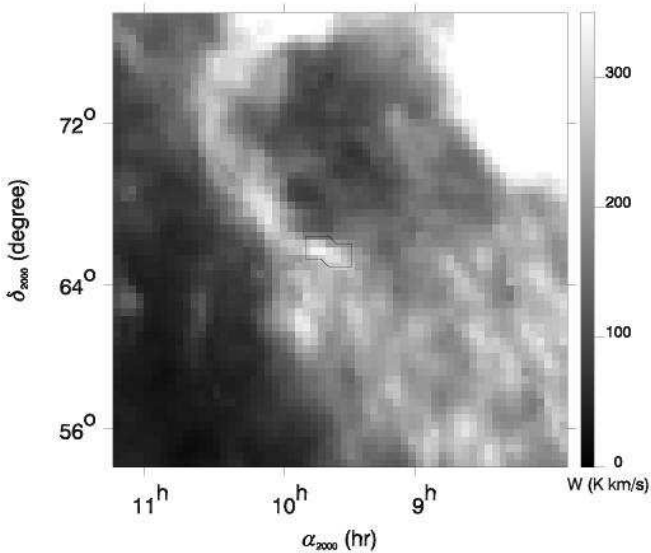}
\caption{\label{ncl_leiden} Integrated emission of the 21 cm Leiden/Dwingeloo survey
in the direction of the North Celestial Loop.
The Ursa Major fields observed with the DRAO interferometer are indicated by boxes.}
\end{figure}

\begin{figure}[!ht]
\includegraphics[width=9cm, draft=false]{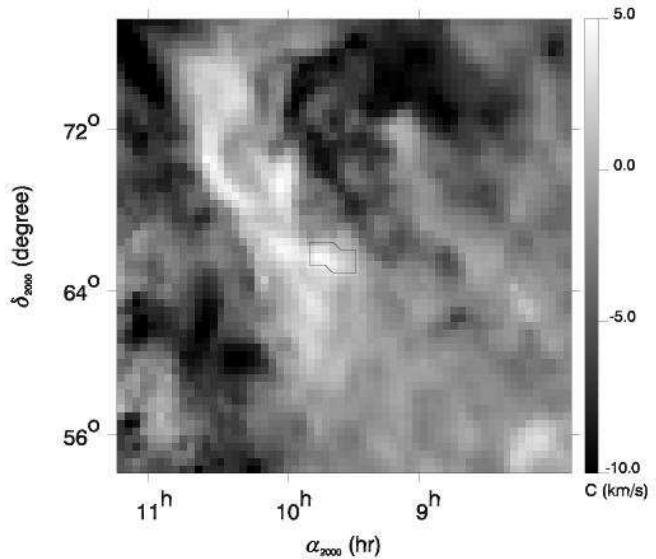}
\caption{\label{velfield_ncl} Centroid velocity field of North Celestial Loop (Leiden/Dwingeloo data).
See Fig.~\ref{ncl_leiden}.}
\end{figure}

\begin{figure}[!ht]
\hspace{-0.7cm}
\includegraphics[width=9cm, draft=false]{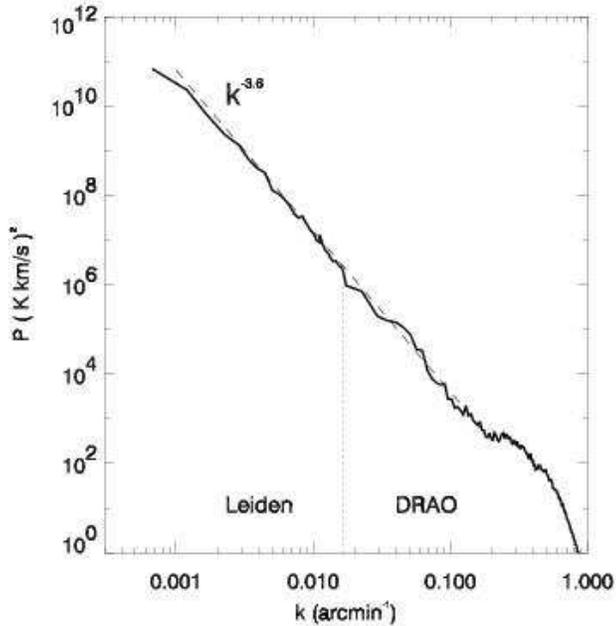}
\caption{\label{ps_ntot_leiden} Power spectrum of the integrated emission of the Ursa Major / North Celestial Loop
region. The dotted line defines the scale regimes of the DRAO and Leiden/Dwingeloo data (see Fig.~\ref{ncl_leiden}).
The dashed line represents a slope of -3.6.}
\end{figure}

\begin{figure}[!ht]
\hspace{-0.7cm}
\includegraphics[width=\linewidth, draft=false]{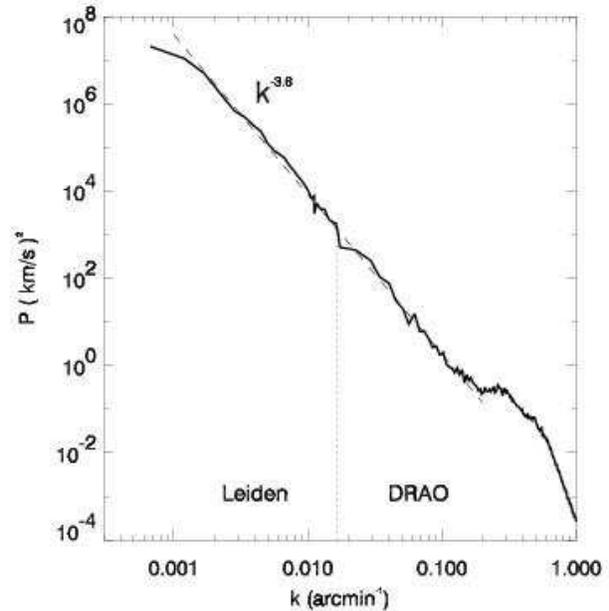}
\caption{\label{ps_velfield_leiden}  Power spectrum of the centroid velocity field of the Ursa Major / North Celestial Loop
region. The dotted line defines the scale regimes of the DRAO and Leiden/Dwingeloo data 
(see Fig.~\ref{velfield_ncl}). The dashed line represents a slope of -3.6.}
\end{figure}

The structure of the centroid velocity field of the DRAO fields shows many similarities 
with the column density map. In both maps, the main filaments of the cirrus are visible, 
indicating that the density and velocity of the \hi show some correlation.
This is also true at larger scales as one can see in Figures~\ref{ncl_leiden} and \ref{velfield_ncl} where
we show the integrated emission and the centroid velocity field of the North Celestial Loop (Leiden/Dwingeloo data). 
At the scale of the loop, the main density structures seen in the integrated emission map are correlated
with structure in the centroid velocity field.

To estimate the correlation level between the column density and the velocity field, we have computed
the cross-correlation defined as:
\begin{equation}
Ccr = \frac{\left< N_{Htot} \, V_c \right> - \left<N_{Htot}\right>\left<V_c \right>}{\sigma_N \sigma_V}
\end{equation}
where $\sigma_N$ and $\sigma_V$ are the standard deviation of the column density and centroid velocity fields
respectively. The cross-correlation is 0.44 for the Ursa Major field and 0.25 for the North Celestial Loop region,
indicating a moderate correlation between the centroid velocity and column density field.

\section{Discussion}

\label{discussion}

As shown by  \cite{lazarian2000} and \cite{miville-deschenes2003b}, 
the interpretation of the power spectrum of individual channel maps is not unique since brightness 
fluctuations in a channel map are the combination of velocity, density and temperature fluctuations. 
On the other hand, for optically thin and isolated regions,
and at scales that are smaller than the depth of the observed region,
the power spectrum of the \hi integrated emission is exactly the power spectrum of the 3D density field.
Furthermore, under these conditions and in the case of Gaussian fields, \cite{miville-deschenes2003b} show that
the power spectrum of the centroid velocity field is exactly the power spectrum of the 3D velocity field.

With these ideas in mind, it is interesting to 
compare our results ($\gamma_n=-3.6\pm0.2$ and $\gamma_v=-3.6\pm0.2$)
with previous power spectrum analysis of 21 cm observations, and try
to draw a coherent picture of the \hi density and velocity statistical properties.
First, \cite{crovisier83,green93,deshpande2000,dickey2001} all found a spectral index of $\sim -2.9\pm0.2$  
for individual channel maps of Galactic plane regions with significant emission, 
in accordance with our values for individual channels in the line center 
(see Figures~\ref{exemple_filtrage} and \ref{expo_vs_channel}). But, as it has been said before, 
the contribution of the velocity, density and temperature
fluctuations to the structure in individual channel maps makes these spectral indexes very difficult to interpret.
Aware of this, \cite{dickey2001} also looked at the spectral index of the integrated emission
which, in their most out-of-plane region ($b=1.3^\circ$), is $\sim -4$. 
However, one must keep in mind that the power spectrum of an integrated emission map 
of Galactic plane regions could be affected by non-turbulent large scale structure and
by significant \hi self-absorption. 
Recent high-resolution (1') 21~cm observations \cite[]{gibson2000} have shown the presence of several
\hi self-absorption regions in the Galactic plane, 
that could not be detected in previous low-resolution 21~cm surveys.
As self-absorption seems to be more important at small-scale, it could reduce significantly the
amplitude of the small-scale \hi emission and produce an artificial steepening of the power spectrum. 
Lets also mention the work of \cite{deshpande2000} who looked at the power spectrum of the velocity-integrated 
\hi optical depth, using absorption measurements in the direction of Cassiopeia~A.
They found a relatively high spectral index ($-2.75\pm0.25$) but, as they pointed out, the optical depth 
depends both on the column density and the gas temperature which makes the relation with the 3D density 
power spectrum not straightforward.

Recently there have also been some power spectrum analysis of 21 cm emission in external galaxies.
For instance \cite{stanimirovic2001} found $\gamma_n=-3.3$ for the \hi integrated emission 
in the Small Magellanic Cloud (scales from 30 pc to 4 kpc). 
In the Large Magellanic Cloud, \cite{elmegreen2001} found that the power spectrum of the \hi integrated emission
is well described by two power laws of indexes -3.66 and -2.66, on spatial scales between 20 pc and 9 kpc. The
spectral index change, which occurs near $L\sim 100$ pc, can be explained by the fact that 
the scales on the plane of the sky are larger than the depth of the
medium for $L > 100$ pc. In this case, the steepening of the slope observed at small scales 
reflects a change in the topology of the medium investigated from a 2D (large scales) to a 3D (small scales) system,
an effect described by \cite{miville-deschenes2003b}.
The density spectral index measured by \cite{elmegreen2001} in the LMC 
is in accordance with our $\gamma_n$ value, even if their smallest scale corresponds approximately to our largest scale.
On the other hand, contrary to \cite{elmegreen2001} we do not observe a break in the slope of the power spectrum
which indicates that the scale range we trace (from 0.1 to 25 pc) is smaller than the typical
depth of the medium. This is in agreement with the fact that the scale height of Galactic \hi is $\sim100$ pc \cite[]{lockman91}.

{ The diversity of \hi spectral indexes found in the litterature 
reveals the importance of gas temperature, depth, energy injection and opacity effects 
on the power spectrum of 21 cm emission and how difficult it is to determine accurately
the 3D statistical properties of interstellar turbulence.
Unlike most of the \hi studies done so far, 21 cm observations of high latitude regions like
the one presented in this work offer the most favorable physical conditions 
to determine the 3D density and velocity power spectrum of interstellar turbulence.
Such regions are almost unaffected by self-absorption
and their density and velocity fields are dominated by turbulent motions,
which is not always the case for Galactic plane regions and external galaxies where 
energy injection at several scales by supernovae, stellar winds or outflows might
affect the power spectrum.
The work we present in this paper is the first study of the inertial range of interstellar
turbulence, far from energy injection and dissipation scales, and where opacity
and projections effects are controled.
It is the first time that the 3D density and velocity power spectra,
essential quantities for the understanding of interstellar turbulence, 
are determined for the Galactic interstellar medium.
Furthermore, the dynamical range of scales over which they are measured (more than two orders
of magnitude) is larger than for all previous studies. The existence of these two scaling laws is therefore
a robust result.}

{It is important to point out that our results ($\gamma_n=-3.6\pm0.2$ 
and $\gamma_v=-3.6\pm0.2$) agree with the Kolmogorov prediction for 
incompressible turbulence (-11/3). It may indicate that compressibility
and magnetic field do not affect significantly the energy cascade 
of interstellar turbulence.
Our results are in accordance with the work of \cite{goldreich95} 
(see also the review of \cite{cho2003}) 
who predicts a Kolmogorov type energy spectrum for an incompressible turbulent 
magnetized fluid. But \cite{goldreich95} also predicts an anisotropy of the 
velocity and density fluctuations that increases at small scales. 
It is interesting to point out here that the density and
velocity structures of the Ursa Major cirrus are clearly anisotropic; filaments
with aspect ratio of 15 are observed. One important step in characterising interstellar
turbulence would be to quantify this anisotropy and its variation with scale.
}

It is also interesting to note that the spectral indexes of the \hi are significantly different
from what is observed with molecular or dust emission. Several studies of integrated CO emission
shows spectral indexes around -2.5 and -2.8. For instance \cite{bensch2001} found relatively
uniform spectral index values ($-2.8 < \beta < -2.5$) for a set of five molecular clouds observed
with tracers of different optical depth ($^{12}$CO and $^{13}$CO $J=1\rightarrow0$).
The work of \cite{gautier92} on 100 \um dust emission also shows a relatively shallow slope 
$\beta \sim -3$ for non-starforming regions. These results seem to indicate more small scale structures
than what is observed in \hip. 
However molecular emission is, contrary to \hip, strongly affected by opacity and heating variations within
the cloud which may significantly modify the power spectrum. Similarly the dust thermal emission 
may vary locally according to the grain properties \cite[]{miville-deschenes2002}, to the
extinction of the stellar radiation or the abundance of H$_2$. This apparent difference between 
the atomic and molecular gas structure clearly deserves further analysis. 

\section{Conclusion}

We have presented a power spectrum analysis of 21 cm interferometric observations
of the Ursa Major / North Celestial Loop region. These observations of an isolated
high-latitude region, with no star-formation activity, allow to study the statistical properties 
of the inertial range cascade of interstellar turbulence in the \hip. 
From the analysis of the integrated emission, of velocity channels and of the centroid
velocity field of the \hi emission 
with a few assumptions on the depth of the emitting medium and its
transparency, we were able to estimate the spectral indexes
of the density and velocity fields in 3-D. For both fields we find a
spectral index of $-3.6\pm0.2$. Last the velocity field and the integrated emission
map show moderate correlation with a value of at most 0.4 in the Ursa Major field.

\section*{Acknowledgments}

{\em The authors thank the team at the DRAO observatory for their 
help in the data reduction and Snezana Stanimirovic for very useful comments.
The Fond FCAR du Qu\'ebec and the National Science and Engineering Research 
Council of Canada provided funds to support this research project.}

\end{document}